\documentclass{aastex}

\input psfig.sty

\def\cgs{{ erg cm$^{-2}$ s$^{-1}$}}
\def\etal{et al.}

\begin{document}
%%%%%%%%%%%%%%%%

\shorttitle{IRON K--LINE IN GRB000214 X--RAY SPECTRUM}
\shortauthors{ANTONELLI ET AL.}

\title{Discovery of a redshifted Iron K--line in the X--ray afterglow of GRB~000214.}

\author{L.A. Antonelli\altaffilmark{1}, 
L. Piro\altaffilmark{2},
M. Vietri\altaffilmark{3}, 
E. Costa\altaffilmark{2}, 
P. Soffitta\altaffilmark{2},  
M. Feroci\altaffilmark{2}, 
L. Amati\altaffilmark{4}, 
F. Frontera\altaffilmark{4,5}, 
E. Pian\altaffilmark{4}, 
J.J.M. in 't Zand\altaffilmark{6}, 
J. Heise\altaffilmark{6}, 
E. Kuulkers\altaffilmark{6}, 
L. Nicastro\altaffilmark{7}, 
R.C. Butler\altaffilmark{4,8},
L. Stella\altaffilmark{1,9},
and G.C. Perola\altaffilmark{3}
}

\altaffiltext{1} {Osservatorio Astronomico di Roma, Via Frascati, 33, 00040, Monteporzio Catone (Roma), Italy.}
\altaffiltext{2} {Istituto di Astrofisica Spaziale, C.N.R., Area della Ricerca di Tor Vergata, Via Fosso del Cavaliere, 100, 00133, Roma, Italy.}
\altaffiltext{3} {Universit\'a degli Studi ``Roma Tre'', Via della Vasca Navale, 84, 00146, Roma, Italy.}
\altaffiltext{4} {Istituto Tecnologie e Studio delle Radiazioni Extraterrestri, C.N.R., Via Gobetti, 101, 40129, Bologna, Italy.}
\altaffiltext{5} {Dipartimento di Fisica, Universit\'a di Ferrara, Via del Paradiso, 12, 52100, Ferrara, Italy.}
\altaffiltext{6} {Space Research Organization Netherlands, Sorbonnelaan, 2, 3584 CA Utrecht, Netherlands.}
\altaffiltext{7} {Istituto di Fisica Cosmica e Applicazioni Calcolo Informatico, C.N.R., Via La Malfa, 153, 90138, Palermo, Italy.}
\altaffiltext{8} {Agenzia Spaziale Italiana, Viale della Regina, 212, 00100, Roma, Italy.}
\altaffiltext{9} {Affiliated to the International Center for Relativistic 
Astrophysics}

\begin{abstract}
We report the detection ($3\sigma$ significance level) of a
strong iron emission line in the X-ray spectrum of the afterglow of
GRB 000214 (``Valentine's Day Burst'') observed by BeppoSAX. An
emission line feature was observed with a centroid energy of
$4.7\pm0.2$ keV which, if interpreted as K$\alpha$ emission from
hydrogen-like iron, corresponds to a redshift of z=0.47. The
intensity (EW$\sim 2$~keV) and duration (tens of hours) of the line
give information on the distance, from the burst region, of the
emitting material ($R\geq3\times10^{15}$ cm) and its mass
($M\geq1.4M_{\odot}$). These results are not easily reconciled with
the binary merger and hypernova models for gamma ray bursts, because 
they require large amounts of mass (about $1 M_\odot$), at large
distances ($\approx 10^{16}\: cm$), and at newtonian speeds. 
  
\end{abstract}

\keywords{gamma rays: bursts---line: formation---X-rays: general}

\section{Introduction}

Three years after the first identification of the X-ray afterglow from
a Gamma Ray Burst (GRB) (Costa et al., 1997), the nature of Gamma Ray
Bursts' progenitors remains uncertain. The most widely discussed
theoretical models for GRBs consider vastly different scenarios both
in terms of the progenitors and environment. If the GRBs originate
from neutron star-neutron star or neutron star-black hole mergers
(Narayan, Paczynski and Piran 1992) then, for about a half of all
GRBs, the explosion should be located far from the place where the
progenitor binary system was formed and most probably in a low-density
interstellar medium. On the other hand, if GRBs originate from
Hypernovae (Paczynski, 1998; Woosley, 1993) or SupraNovae (Vietri \&
Stella, 1998), the immediate progenitor is a massive, rotating star,
and thus the explosion should take place in a high density medium,
probably a star-forming region. The discovery of emission
%of strong and possibly variable (with respect to the continuum) 
lines in the X-ray spectrum of a GRB afterglow represents a major step
toward the understanding of the nature of GRB progenitors. In fact,
the measurement of X-ray lines emitted by a GRB or its afterglow 
provides a direct measurement of the redshift and a powerful diagnostic
of both the nature of the central engine and the environment in which
GRBs go off (Perna \& Loeb, 1998; M\`esz\`aros \& Rees, 1998;
B\"ottcher et al., 1999, Ghisellini et al., 1999; Paerels et al.,
2000). Other methods, such as collecting offsets between GRBs
afterglow and their host galaxies (Bloom et al., 1999) or searching
for absorbing material between the GRB and the observer (Wijers \&
Galama, 1999; Owens et al., 1999), are comparatively less direct.

Presently, there are only
two marginal detections of an X-ray emission line in GRB
afterglows,  one in GRB 970508 (chance
probability for a statistical fluctuation of 0.7\%, Piro et al. 1999)
and one in GRB 970828 (probability of 1.7\%, Yoshida et al.
1999).  The presence of a strong iron line in the X-ray spectrum of a
GRB afterglow implies a rich environment located very close to the GRB
region and may be an important clue in favor of collapsar models
(M\`esz\`aros \& Rees, 1998; Lazzati, Campana, \& Ghisellini, 1999;
Weth et al., 2000; Vietri et al., 1999). In fact, in the case of the
Hypernova scenario an iron rich circumburst environment may be
produced by the stellar wind before the explosion of the Hypernova
(M\`esz\`aros \& Rees, 1998). A similarly favorable situation 
occurs in the SupraNova scenario (Vietri \& Stella, 1998) where a
supernova explosion precedes by a few months the GRB event.
GRB 000214 (``Valentine's Day Burst'') has no
optical nor near-infrared afterglow (Rhoads et al., 2000;
Wijers et al., in preparation) as is the case for about half of all
GRBs with follow--up observations (e.g., Kulkarni et al., 2000).  The 
lack of optical
and IR counterparts to a GRB may result from extinction in a
dense surrounding medium; therefore, searching for emission lines in
the X-ray afterglows of these GRBs looks especially promising.  Based
on a BeppoSAX pointed observation of GRB 000214, we report on
the detection of an Iron K$\alpha$ line from its X-ray afterglow.

\section{Observation and results}

GRB 000214 was detected both by the Gamma Ray Burst Monitor (GRBM) 
and Wide Field Cameras (WFC) on-board BeppoSAX on 2000 Feb. 14 01:01:01
UT.  The prompt event, i.e. the gamma ray emission detected by GRBM,
lasted $\sim 10$~s and had a fluence of $\cal F$=$1.4\times10^{-5}$
erg cm$^{-2}$ in the 40--700 keV band. The corresponding prompt X-ray
emission detected by WFC had a fluence of $\cal F$=$1.0\times10^{-6}$
erg cm$^{-2}$ (2--10 keV). In the X-rays a complex multi-peaked
structure lasting about 2 min after the burst proper was clearly seen;
a detailed analysis of the early behaviour of GRB 000214 will be
presented elsewhere.  The evolution of the X to Gamma-ray spectrum
from the prompt event showed a spectral softening similar to that
observed in other GRBs (Frontera et al., 1998, 2000a, 2000b).

\subsection{Narrow Field Instruments observation}

A follow-up observation with the BeppoSAX Narrow Field Instruments
(NFI) began about 12 hours after the burst and lasted about 104,000
s. The effective exposure time was 51,000 s on source time for the
BeppoSAX Medium Energy Concetrator/Spectrometer (MECS), and 15,000 s
in the Low Energy Concentrator/Spectrometer (LECS). A previously
unknown X-ray source, 1SAX J1854.4-6627, was detected in the MECS and
LECS field of view at a position of RA =$18^{h}~54^{m}~27^{s}.0$ and
DEC = $-66^{o}~27'~30''$ (J2000) (uncertainty radius of 50'').  The
source position lies within the WFC error circle (Paolino et al.,
2000; Antonelli, 2000) and IPN error box (Hurley \& Feroci, 2000).
The 2--10 keV flux of 1SAX J1854.4-6627 decreased by a factor of about
two, from $(7.7\pm0.8)\times10^{-13}$ (first 11 ks) to
$(4.2\pm0.7)\times10^{-13}$ \cgs (last 10 ks), during the BeppoSAX
observation (these fluxes were derived by using the best fit model to
the spectra, see next section).  In consideration of its fading
behaviour and position within both the WFC and IPN error boxes, we
concluded that 1SAX J1854.4-6627 represents the afterglow of
GRB~000214.

The MECS lightcurve yields a flux decay ($F(t)\propto t^{-\beta}$) with
$\beta=(0.8\pm0.5)$ (hereafter all errors are quoted at
90\% confidence level for one parameter of interest). Combining the
MECS lightcurve with the WFC lightcurve from after the main event
(extending for 39 s after burst onset) yields a flux decay with a
power-law slope of $\beta=(1.41\pm0.03)$.  This slope is somewhat
steeper than usually observed in GRB X-ray afterglows
($\beta\simeq1.2$). 

\subsection{Spectral analysis of GRB 000214 X-ray afterglow}

We extracted the LECS(0.1-4.0 keV) and MECS(1.6-10.0 keV) spectra
of the source from the entire BeppoSAX observation (figure 1) using
standard procedures (Fiore et al., 1999)\footnote{Fiore, F., Guainazzi,
M., \& Grandi, P., 1999, Cookbook for BeppoSAX NFI Spectral Analysis,
{\it (http://www.sdc.asi.it/software/)}}.  The extraction regions were
chosen so as to maximize the signal-to-noise ratio.
%These were 2 and 3 arcmin radius circles, for the MECS and LECS, 
%respectively.
An accurate analysis of local background was performed in order to
check for spurious features. After having verified that the local
background is compatible with that derived from blank field
observations available at the BeppoSAX Science Data Center, we used
the latter in order to achieve a background subtraction with improved
signal to noise (see also figure 2).  Both the LECS and MECS spectra
were binned so as to have at least 15 photons per bin. Spectral
analysis was performed using the XSPEC10.0 package. We checked that
our results are nearly insensitive to the details of the background
subtraction.

We first fit the data with a power-law plus photoelectric absorption. The
relative normalization of LECS versus MECS was left free to vary in
the 0.7-1.0 range (Fiore \etal, 1999).  We obtained a best-fit
value of $\chi^{2}=25.4$ for 14 (d.o.f.). The column density
best-fit value ($N_{H}=0.7^{+7.5}_{-0.7}\times10^{20}$ cm$^{-2}$) is
compatible with the Galactic value $N_{H}=5.5\times10^{20}$ cm$^{-2}$
(Dickey \& Lockman, 1990), so we fixed it to this value.  We obtained
a power-law photon index of $\Gamma=2.0\pm0.3$ and a flux of
$F_{x}(2-10~keV)=(2.75\pm0.9)\times10^{-13}$ \cgs.
%normalization of $(1.2\pm0.4)\times10^{-4}$~photons
%keV$^{-1}$~cm$^{-2}$~s$^{-1}$ at 1~keV.  
The best-fit value was $\chi^{2}=27.5$ for 15
(d.o.f.). These relatively high chisquare values (corresponding to a
chance probability of $P_{\chi}(\chi^{2},\nu)\simeq0.02$ in the latter
case) justified a deeper analysis. The residuals to
the fit above show a clear excess around energies of $\sim 4-5$~keV
which can be interpreted as an emission line (see figure 1). We added
a narrow (with respect to MECS energy resolution, which is
$\Delta$E/E=8$\times$(E/6 keV)$^{-0.5}$ FWHM\%) Gaussian line to the
previous model in order to fit the observed feature.  We inferred in
this way a line centroid energy of $E=4.7\pm0.2$ keV and an intensity
of $I_{l}=(9\pm3)\times10^{-6}$ photons cm$^{-2}$ s$^{-1}$,
translating into an E.W. of $\sim2.1$~keV. The best fit chisquare
value was $\chi^{2}=11.5$ for 13 d.o.f. Applying an $F$--test to
assess the statistical significance of the Gaussian feature we
obtained a chance probability of 0.27\% (corresponding to $3.0\sigma$
significance level). The other spectral parameters were only
marginally affected by the inclusion of the Gaussian, with a power-law
photon index of $\Gamma=2.2\pm0.3$ and a flux of
$F_{x}(2-10~keV)=(2.9\pm0.9)\times10^{-13}$ \cgs.
%a normalization $(1.3\pm0.4)\times10^{-4}$
%photons keV$^{-1}$ cm$^{-2}$ s$^{-1}$ at 1 keV. 
The LECS normalization relative to the MECS depends mainly on the
source position in the LECS and the different exposure intervals than
the MECS (note that the source was fading). However the statistical
uncertainty we determined for this normalization was greater than the
nominal range; therefore we also fixed it to 0.8, the value inferred
from a BeppoSAX observation of NGC 7469 (Piro \etal 1999b), a Seyfert
1 galaxy with a $0.1-10$~keV spectrum similar to that of the GRB
000214 afterglow.  The best fit chisquare value was $\chi^{2}=28$ for
16 d.o.f. fitting data with the simple power-law model and
$\chi^{2}=12.1$ for 14 d.o.f. adding a Gaussian line to the previous
model. Following this procedure, while the spectral parameters
remain essentially unchanged a slightly improved statistical
significance ($F$--test) was obtained for the fit with the power--law
and Gaussian line (chance probability of 0.15\%, or $\sim 3.2\sigma$).

We divided the MECS observation in two consecutive intervals (20 ks
and 30 ks exposure). MECS spectra were extracted from each interval by
the same method as above. As shown in figure 3 both
spectra show evidence for an emission line at an energy compatible, 
within the errors, with the line detected during the entire
observation.  Note also that most of the flux decrease in the second
part of the observation seems due to the continuum rather than
the line. Poor statistics does not allow a more quantitative
investigation.\\

\section{Discussion and conclusions}

The only cosmologically abundant element that can produce lines beyond
4 keV is iron, provided one precludes Doppler shifts to shorter
wavelengths.  Iron emits a K$\alpha$ line photons with an energy
between 6.4 (for the lower ionization stages of iron) and 6.95 keV
(for hydrogen-like iron). The observed feature may be also interpreted
as the Iron recombination K--edge at 9.28 keV (corresponding to a
redshift of $z=0.9$) but, in such a case, according to Weth et
al. (2000), we should expect both a Fe K--edge and a K$\alpha$ line
having about the same intensity. The Fe-K$\alpha$ line should be
redshifted at an energy between 3.2 and 3.5 keV but no line is
detected in the MECS spectrum of GRB 000214 afterglow within this
range.  If we identify the emission feature in GRB 000214 as Fe
K$\alpha$, the corresponding redshift is between $z=0.37$ and 0.47. We
will assume $z=0.47$ for simplicity, but none of the following discussion
hinges upon this exact value.

The case of GRB 000214 differs from that of GRB 970508 in two
ways. First, the iron line is more significant in GRB 000214: 
we find that the probability of a chance fluctuation is $0.27\%$, as
compared to $0.7\%$. This difference becomes even more remarkable when
one considers that there is no independent determination of the
redshift for GRB 000214, while GRB 970508 has $z = 0.835$ from optical
observations (Metzger et al., 1997). 
Second, the afterglow observation of GRB 000214 did not
allow detection of a possible reburst, contrary to GRB 970508. In
fact, GRB 000214 was observed just once after the burst proper,
beginning $12$ h after the main event, and for a total of $\approx
10^5$ s.  GRB 970508, instead, was observed on four different
occasions after the burst, beginning $6$ h after the main event, with
the last set of observations more than $6$ days after the burst (Piro
et al., 1999). Thus, for GRB 000214, we have no time coverage to
exclude a reburst before or after the time of our observations: all we
can say is that, during this limited time, the flux
decreased by about a factor of $2$ (see Section 2.1).  This is
consistent with GRB 970508, where the line was bright as the
flux was decreasing (Fig. 2a of Piro et al., 1999); thus the model
of Vietri et al. (1999), connecting the appearance of the iron line to
the reburst, is still viable.

For $z = 0.47$, $H_\circ = 70\;$ km s$^{-1}$
Mpc$^{-1}$, and $\Omega = 1$, we find a total energy release, in
the line only, of $E_l = 3\times 10^{48}\;$ erg, i.e. $N_l
= 3\times 10^{56}$ photons emitted within the first $t_d \approx 10^5$ 
s after the
burst; the emission is isotropic because the line is
sub--relativistic. If each iron atom were to emit a single line
photon, then $M_{Fe}\approx 12 M_\odot$, way too large. 
Thus each iron atom must produce several line photons (Lazzati \etal,
1999), and, consequently, the time--scale $t_{Fe}$ over which
every iron atom produces a photon satisfies $t_{Fe} \ll t_{d} \approx
10^5\;$ s, and does not seem to begin fading. This implies that the
time--scale $t_d$ is set by the geometrical factor: the line--emitting
material must be located at $R \ga 3\times 10^{15}\;$ cm, and we are
seeing some sizeable fraction of it.

The minimum iron mass is determined by Lazzati \etal (1999) as
\begin{equation}
M_{Fe} \geq 5\times 10^{-3} M_\odot \frac{F_{-13}t_5 R_{16}^2}{q E_{50}}
\end{equation}
where $F_{-13} \approx 0.5$ is the line luminosity in units of
$10^{-13}$ \cgs, $t_5 \approx 1$ is the line duration in units of
$10^5\;$ s, $R_{16}\approx 0.3$ the material distance in units of
$10^{16}\;$ cm, $q$ is the fraction of the absorbed ionizing fluence and
reprocessed into the line ($q\la 0.1$ as derived by Ghisellini et al.,
1999) and $E_{50}$ is the total X--ray burst luminosity 
in units of $10^{50}$ erg; for GRB 000214 we find $ E_{50}
\approx 8$ if the burst emission were isotropic. With these values 
\begin{equation}
\label{iron}
M_{Fe} \ga 2.4\times 10^{-3} M_\odot \;.
\end{equation}

This argues against a possible origin of this GRB in a neutron
star - neutron star (or neutron star - black hole) binary merger. The
problem is not one of total iron content of the model: for realistic
parameters, this scenario can easily satisfy the above limit. The
problem lies instead in the capacity of the model to deposit this much
iron at these large distances, and at Newtonian speeds.  In fact,
under our assumption that the line is not blueshifted (at least, its 
Lorenz factor is
$\approx 1$), the iron cannot have been released in the burst event
itself: there is not time enough for matter ejected in the explosion
to rush ahead of photons, slow down, and then be hit by the burst
flash. The material generating the line must have been released well
before the burst. Given their lack of atmospheres and winds, the only
mechanism capable of releasing matter from inspiralling neutron stars
is tidal interaction. A mass loss estimate for this process by
M\`esz\`aros and Rees (1992) has a lower limit of $M_{th} = 
4\times 10^{30}\; g$,
enough to account (barely!) for Eq. \ref{iron}, but this optimistically 
assumes that the typical viscosity damping time-scale equals the light travel
time across the star (Bildsten and Cutler 1992). Though the exact amount of 
viscosity in a neutron star is not well--known, theoretical arguments 
fall short of the amount required for the correctness of the estimate 
by M\`esz\`aros and Rees (1992), by about a factor $2000$ 
(Kochanek 1992, Eq. 4.7, Bildsten and Cutler 1992, see the discussions after 
Eq. 9 and 16). 

In hypernova models, an intense wind should characterize the
massive progenitor before the collapse. Calling $A$
the iron mass fraction relative to the solar abundance (Lang 1991, p. 84), 
we expect from Eq. 2  a minimum mass of
\begin{equation}
M \geq 1.4 M_\odot \frac{1}{A E_{50}} \;. 
\end{equation}
The radial mass distribution from a massive stellar wind is
$d\!M = \dot{M} d\!r/v_\infty$.  The most extreme ratios of stellar
wind mass loss rate $\dot{M}$ to asymptotic speed $v_\infty$ occur in
Luminous Blue Variables and Red SuperGiant
Variables ($\dot{M} = 10^{-4} M_\odot\;$ yr$^{-1}$, $v_\infty = 10-100$ km
s$^{-1}$, Lamers, 1995, van Loon et al., 1999). For these winds, at
$R\approx3\times 10^{15}$~cm there can only be $M\approx
10^{-2}-10^{-3} M_\odot$, missing the above lower limit by $2-3$
orders of magnitude. The densest circumstellar envelopes 
around the progenitors of some peculiar core collapse
supernovae ({\it e.g.}  $n\sim10^8-10^9$~cm$^{-3}$ at $R\sim
3\times10^{15}$~cm in SN1997ab, Salamanca et al., 1998) imply a mass
of $M\sim10^{-1}-10^{-2}~M_{\odot}$, substantially
lower than required by Eq. 3.  A model immune from
these inconsistencies is the SupraNova, where a core collapse
supernova explosion precedes the GRB event by months to years (Vietri and
Stella 1998). 

On the other hand, the lack of large amounts of material along the
line of sight, as testified by the absence of photoelectric soft X-ray
absorption on top of the Galactic value\footnote{The upper limit
derived above, $N_H = 2.7\times 10^{20}$ cm$^{-2}$, for the
non--Galactic column density, corresponds to a total mass, within $12$
light hours, of $5\times 10^{-6} M_\odot$, which could indicate a
large amount of mass only for implausibly large ($\approx
1-10^{-6}$) ionization fractions.}, and the development of a normal
X--ray afterglow, clearly suggest the presence of outlying material
located in the progenitor's equatorial plane, at large angles from the
line of sight. This material may be either left over from the
progenitor's formation process (as in a primordial gas cloud) or
spewed out by the progenitor. Key future observations capable of 
discriminating between these possibilities will be the measurement of 
the line width, and a search of atomic species produced in the interiors 
of evolved stars. Velocities of $\ga 1000\; km\; s^{-1}$ can rule 
out any primordial cloud and/or hypernovae winds; the detection of
cobalt or nickel may establish the recent origin of the material from
a SN explosion.

\medskip

This research is supported by the Italian Space Agency (ASI) and
Consiglio Nazionale Ricerche of Italy. Beppo-SAX is a joint program of
ASI and the Netherlands Agency for Aerospace Programs (NIVR).  We 
thank all members of the Beppo-SAX SDC, SOC and OCC. 
%Science Data Center,
%Scientific Operation Center and the Operation Control Center. 
We also thank L. Burderi, G.L. Israel, F. Fiore, G.Matt and G. Bono and an 
anonymous referee for helpful comments. 

%Correspondence should be
%addressed to L.A.A. (e$-$mail: angelo.antonelli@coma.mporzio.astro.it).

%%%%%%%%%

\newpage

\begin{figure}
\centerline{\psfig{figure=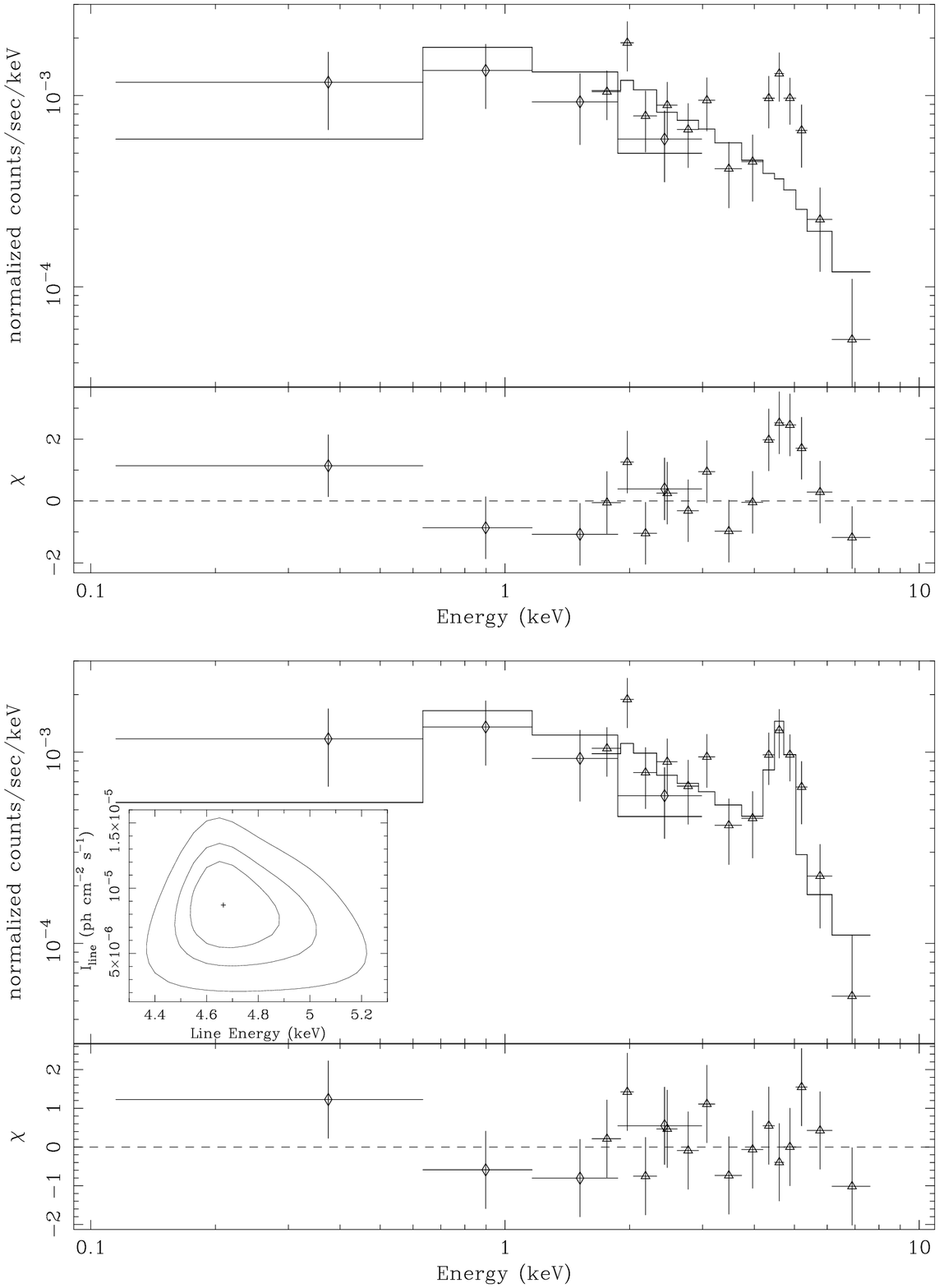,height=18cm} }
\caption{ BeppoSAX (0.1--10.0 keV) spectra of GRB 000214 X-ray
afterglow.  LECS points (0.1--3.0 keV): open
diamonds, MECS points (1.6 -- 10.0 keV): open
triangles.  {\it Top panel:} LECS+MECS spectra fitted with an absorbed
power-law; an excess around 4.7 keV is clearly seen in the
residuals.  {\it Bottom panel:} LECS+MECS spectra fitted with an
absorbed power-law plus a narrow Gaussian line. The inset 
shows the contour plot of the line intensity vs energy. Contours 
correspond to 68\%, 90\% and 99\% confidence levels for two interesting
parameters.}
\end{figure}

\begin{figure}
\centerline{\psfig{figure=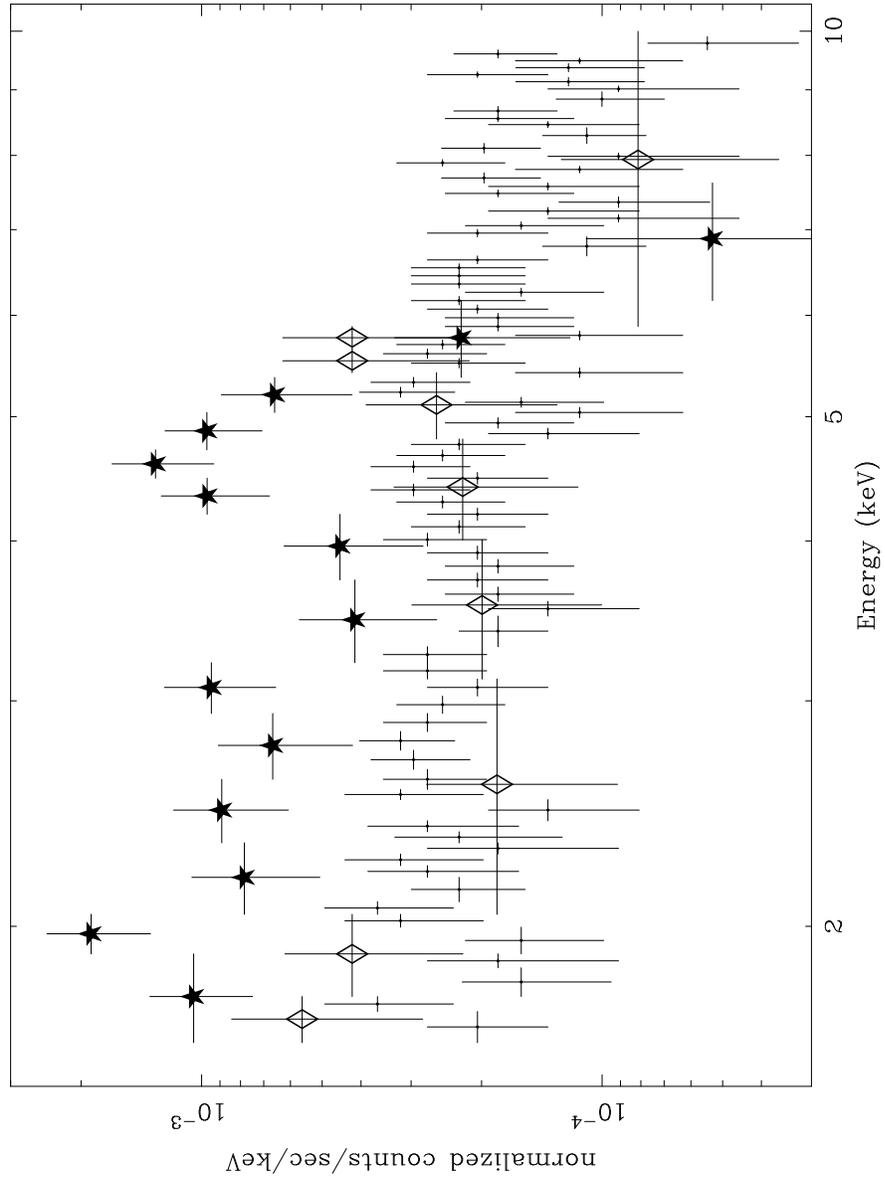,height=18cm} }
\caption{Background-subtracted MECS spectrum of GRB 000214 X-ray afterglow 
(filled stars) compared to the local background (open diamonds) and 
library background (dots) spectra. The feature centered around 4-5 keV is 
about a factor of five higher than the background. }
\end{figure}

\begin{figure}
\centerline{\psfig{figure=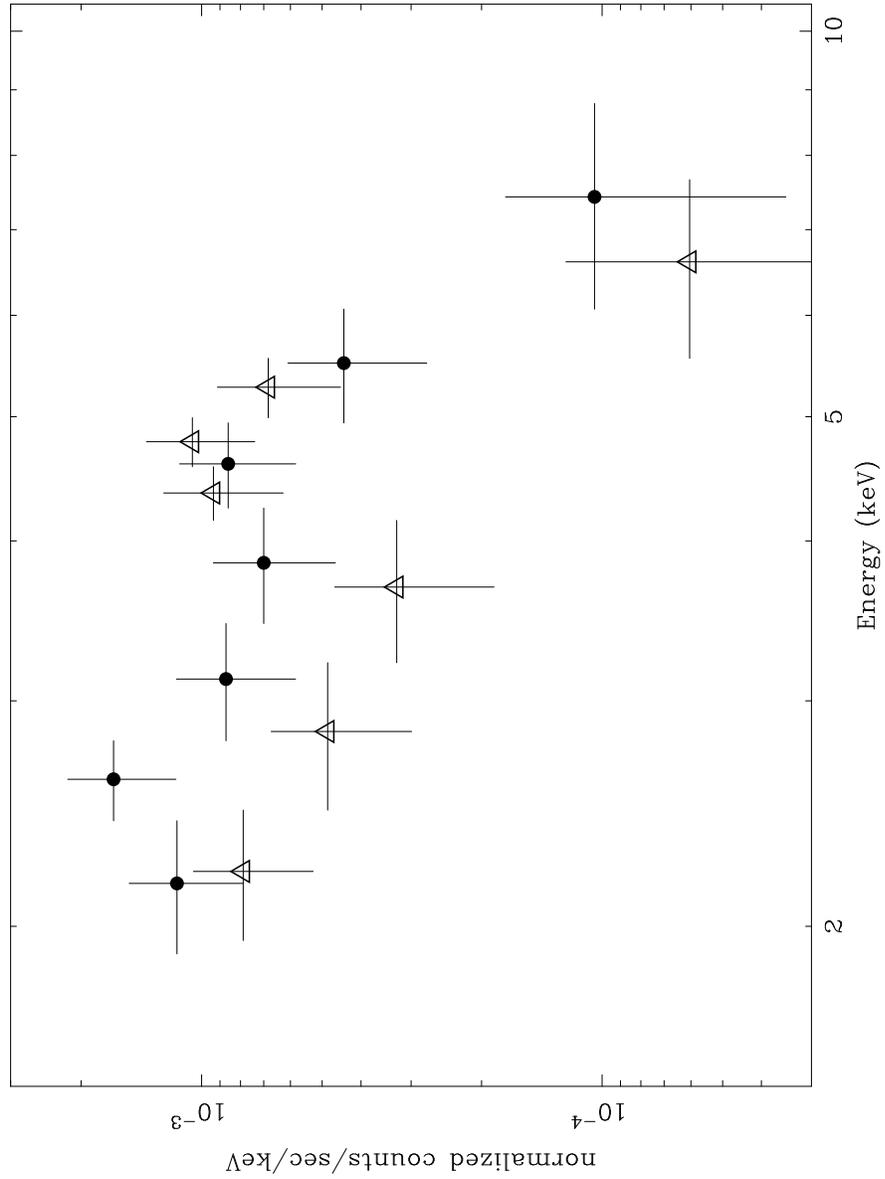,height=18cm} }
\caption{2--10 keV MECS spectrum from the first 20 ks of 
the observation (filled dots) compared to the spectrum from  
the last 30 ks (open triangles). 
Despite poor statistics, 
there is indication that in the second part of the observation 
the continuum faded by a factor of about two, while the excess 
around 4-5 keV remained constant.}
\end{figure}

\end{document}